\documentclass[preprint]{aastex}
\usepackage{graphicx}
\usepackage{txfonts}
\usepackage{natbib}
\usepackage{longtable} 

\def\xte{XTE\,J1810-197}

\def\uu {4U\,0142$+$614}
\def\ee {1E 1048.1$-$5937}

\def\rxj {1RXS J170849$-$400910}
\def\ea {1E 2259$+$586}

\def\src{CXOU J164710.2$-$455216}
\def\wes{Westerlund 1}
\def\swift{{\em Swift}}
\def\XMM{{\em XMM-Newton}}
\def\chandra{{\em Chandra}}
\def\ergscm2{\rm erg\,cm^{-2}\,s^{-1}}
\def\ergs{\rm erg\,s^{-1}}

\newcommand {\rc}{\rm}

\slugcomment{Submitted: 2006 November 20 / Accepted: 15 March 2007 }

\shorttitle{The awakening of CXOU J164710.2--455216}
\shortauthors{ISRAEL ET AL.}

\begin{document}


\title{ \LARGE {\sc The Post-Burst Awakening 
of the Anomalous X-ray Pulsar in Westerlund 1}}


\author{G.L. Israel\altaffilmark{1}, 
S. Campana\altaffilmark{2}, S. Dall'Osso\altaffilmark{1}, M. P.
Muno\altaffilmark{3}, J. Cummings\altaffilmark{4},  R. Perna\altaffilmark{5}, 
and L. Stella\altaffilmark{1}}

\altaffiltext{1}{INAF -- Osservatorio Astronomico di Roma, Via Frascati 33, 
       I--00040 Monteporzio Catone (Roma),  Italy} 

\altaffiltext{2}{INAF -- Osservatorio Astronomico di Brera, Via Bianchi
	46, I--23807 Merate (Lc), Italy}	

\altaffiltext{3}{Space Radiation Laboratory, California Institute of Technology,
Pasadena, CA 91125, USA}


\altaffiltext{4}{University of Maryland, Baltimore County / NASA Goddard Space
Flight Center, Greenbelt, MD 20771, USA}


\altaffiltext{5}{JILA, Univ. of Colorado, Boulder, CO 80309-0440, USA}


\begin{abstract}
On September 21, 2006, an intense ($\sim$10$^{39}\;\ergs$) and short
(20 ms) burst was detected by \swift\ BAT at a position consistent
with that of the candidate Anomalous X--ray Pulsar \src, discovered by
\chandra\ in 2005.  \swift\ follow-up observations began $\sim$13
hours after the event and found the source at a 1-10\,keV flux level
of about 4.5$\times$10$^{-11}\ergscm2$, i.e. $\sim$300 times brighter
than measured 5 days earlier by \XMM. 
We report the results obtained from \swift\ BAT observations of the 
burst and subsequent \swift\ XRT observations carried out during the
first four months after the burst. These data are complemented 
with those from two \XMM\ observations (carried out
just before and after the BAT event) and four archival \chandra\ 
observations carried out between 2005 and 2007.   We find a phase coherent
solution for the source pulsations  after the burst. The evolution of the pulse
phase  comprises an exponential component decaying  with timescale of 1.4\,d 
which we interpret as the recovery stage following a large glitch ($\Delta \nu /
\nu \sim 6 \times 10^{-5}$).  We also detect a quadratic component corresponding
to a spin-down rate of $\dot{\rm P}\sim 9 \times 10^{-13}$ s s$^{-1}$, implying
a magnetic field strength of 10$^{14}$ Gauss.

During the first \swift XRT observation taken 0.6 days after the burst, the
spectrum showed a $kT\sim0.65$\,keV blackbody ($R_{BB}\sim1.5$\,km) plus  
a $\Gamma\sim2.3$ power-law accounting for about 60\% of the 1-10\,keV
observed flux.  Analysis of  \chandra\ archival data, taken during 2005 when the
source was in quiescence,  reveal that the modulation in quiescence is 100\%
pulsed at energies above $\sim$4 keV and consistent with the (unusually
small-sized) blackbody component being occulted by the neutron star as it
rotates.  

These findings  demonstrate that \src\ is indeed an AXP; we compare them 
with the properties of three other AXPs which displayed similar
behavior in the past.
\end{abstract}





\section{\Large {\sc Introduction}}
In recent  years there has been a large observational and theoretical effort
aimed at unveiling the nature of a sample of peculiar high-energy pulsars,
namely the Anomalous X--ray Pulsars (AXPs; 8 objects plus one candidates)
and the
Soft $\gamma$-ray Repeaters (SGRs; 4 objects plus {\rc three} candidates).
SGRs were discovered in  the seventies through the very intense bursts
that they sporadically emit in the soft $\gamma$-ray band; AXPs  were
recognized as a distinct class of X--ray pulsars only a decade ago
by virtue of their peculiar persistent emission and spin--down properties
in the X--ray band ({\rc Paczy\'nski 1992;} Mereghetti \& Stella 1995; for a
recent review see Woods \&
Thompson 2004). It is now commonly believed that AXPs and SGRs are linked at
some level, owing to their  similar timing properties (spin periods in the
5-12\,s range and  period derivatives \.P in the  10$^{-11}$--10$^{-13}$
s\,s$^{-1}$ range).
Both classes  have been proposed to
host neutron stars whose emission is powered by the decay of their extremely
strong inner magnetic fields  ($>10^{15}$\,G; Duncan \& Thompson 1992;
Thompson \& Duncan 1995). The detection of X--ray bursts from \ee,
\ea, \uu\ and \xte\ has strengthened the possible connection
between  AXPs and SGRs (Gavriil et al. 2002; Kaspi et al. 2003; Woods
et al. 2005; Kaspi et al. 2006), as well as the magnetar scenario. 

Different types of  X-ray flux variability have been displayed by AXPs,
including
slow and moderate flux changes (up to a factor of a few) on timescales of
years (virtually all the objects of the class),  moderate-intense outbursts
(flux variations of a factor up to 10) lasting for 1-3 years (\ea,
and \ee), and dramatic and intense SGR-like burst activity (fluence of
$10^{36}-10^{37}$ ergs) on sub-second timescales (\uu, \xte, \ea\ and  \ee; see
Kaspi 2006 for a recent review on the X-ray variability).  Particularly
important was the 2002 bursting/outbursting event detected from \ea, the only 
known event in which a factor of $\sim$10  persistent flux enhancement in an AXP
was detected in coincidence of a burst-active  phase during which
the source displayed more than 80 short bursts (Gavriil, Kaspi \& Woods 2003;
Woods et al. 2004). The timing and spectral properties of the sources changed
significantly during this phase, and recovered  within a few days.
The short recovery time is likely  because of the relatively high X-ray
luminosity level of the pre-outburst phase  ($\sim 10^{35}\ergs$ in the
1-10\,keV band). In 2003  The first transient AXP, \xte, was discovered. This
source displayed a factor of $\sim$100  persistent flux enhancement with respect
to the  pre-outburst quiescent luminosity level ($\sim 10^{33}\ergs$) where no 
pulsations were detected. Unfortunately,  since the initial phases of the
outburst were missed, we do not know whether an active bursting phase, similar
to that of \ea,  set in also in this source, leaving several questions
concerning the  mechanisms of the outburst onset  unanswered (Ibrahim et al.
2004; Gotthelf et al. 2004; Israel et al. 2004; Rea et al. 2004).  Until now, no
transient bursting-outbursting  AXP was known.

On September 21st, 2006, a burst was detected by the \swift\ Burst
alert Telescope (BAT) at a position consistent with the AXP candidate
\src\ in the open cluster Westerlund 1 (Krimm et al 2006; Muno et
al. 2006).  The 20\,ms duration of the burst suggested that the origin
of the burst was indeed the candidate AXP. Unfortunately, since the
burst was initially attributed to a nearby Galactic source, the burst
BAT position was not promptly re-observed by \swift.  Moreover, because of 
the relatively low significance of the burst detection, the
event was initially tagged as ``not real". A subsequent careful
analysis of the data confirmed that the detection was indeed real (Krimm et al.
2006). We activated a ToO observation program with \swift\ in order to look for
burst-induced persistent flux variations of the source. The first
\swift\ pointing was carried out 13 hours after the burst. The AXP was
detected at a flux level about 300 times higher than  that  in the previous
measurement (5 days before, from an \XMM\ Guest Observer Program
pointing; Muno et al.  2006b; 1-10\,keV flux level of
$\sim$1.5$\times$10$^{-13}\ergscm2$), hence confirming the transient behavior
of \src\ (Campana \& Israel 2006; Israel \& Campana 2006; Ibrahim et
al. 2004).  

A first radio observation of \src\ was carried out from Parkes
at 1.4\,GHz a week after the outburst onset, with the goal of 
searching for pulsed emission similar to  the case of \xte\ (Camilo et al.
2006). The data put a tight upper limit of 40$\mu Jy$ on the presence
of pulsed emission from the source (Burgay et al. 2006).
{\rc The position of \src\ was also observed serendipitously by IBIS/ISGRI on
board INTEGRAL on 2006  September 22nd  (23ks of effective exposure time): the
source was not detected and  3 sigma upper limits of 5$\times$10$^{-11}
\ergscm2$ in the 20-40 keV band, and 1.7$\times$10$^{-10}\ergscm2$ in the
40-200 keV band  were derived (G\"otz et al. 2006). Near-IR observations were
carried out on 2006 September 29 in the Ks band and reaching a limiting
magnitude of 20.3 (3$\sigma$; Wang et al. 2006): no IR counterpart or variable
object was found at the \chandra\ position of \src. }

In this paper we report the results from the initial \swift\ BAT detection 
and the subsequent \swift\ XRT monitoring campaign consisting of 15 ToO 
observations. We also report on  a re-analysis of archival \chandra\ datasets
of \wes, which allowed us to carry out the first pulse-phase spectroscopic study
of \src\ in quiescence.  Finally, we compare  the observed behavior of \src\
with that of three other AXPs (\xte, \ee\ and \ea), which displayed a similar
phenomenology in the past. 

\section{\Large {\sc Observations and Results}}
The \swift\ data (see Table\,1) were reduced with the standard BAT
and XRT analysis  software distributed within {\tt Heasoft} (v6.0.5) to produce
cleaned event lists.  For the XRT data we considered Windowed Timing
(WT) and Photon Counting (PC)  mode data (see Hill et al.\ 2004 for a full
description of read-out modes)  and further selected XRT grades 0--12 and 0--2
for WT and PC data,  respectively (according to \swift\ nomenclature;
Hill et al. 2004).  XMMSAS v20060628$\_$1801-7.0 and CIAO v3.3.0.1 were
used to reduce the \XMM\ and  \chandra\ data, respectively. Timing and spectral
analyses were carried out with XSPEC v12.2.1, Xronos v5.21, and 
ad-hoc developed pipelines (Israel \& Stella 1996; Dall'Osso et al. 2003).
     
        \subsection{\large {\sc The BAT event}}
The BAT event took place at 01:34:52 in {\rc Barycentric Dynamical Time (TBC)}
on 2006 September 21.
It was the first burst ever detected from an AXP at energies above 20-30\,keV. 
We analyzed the $\sim$20\,ms-integrated 
spectrum by considering the data in the 15--150\,keV energy  range and 
applying an energy-dependent systematic error
vector\footnote{heasarc.gsfc.nasa.gov/docs/swift/analysis/bat\_digest.html}. 
Owing to poor statistics, a preferred model could not be  singled out:
both a blackbody and a power-law component gave $\chi^2_{\nu} \approx$1
(3 degree of freedom, hereafter dof; see Figure 1). A $kT$ of
9.9$\pm^{2.8}_{2.2}$ keV and a $\Gamma$ of 1.8(5) were obtained for the two
models (errors are at 90\%  confidence level). In both cases a fluence was found
of $\approx   10^{-8}$\,erg\,cm$^{-2}$, corresponding to a total energy of 
$\sim 2 \times 10^{37}$\,ergs (for an assumed distance of 5\,kpc).  Assuming an
exponential decay for the burst flux,  we  determined a decay time $\tau$
of 3(1)\,ms (1$\sigma$ confidence level)  from  the 5\,ms binned BAT
lightcurve centered around the trigger (see inset in Figure 1). Compared with
the properties of previously detected AXP bursts, the  event from \src\ 
had a duration within 1$\sigma$  from the log-normal distribution average value
inferred for \ea, while the fluence is significantly larger (by a factor of
$\sim 50$) than the mean (Gavriil et al. 2004).  Out of $\sim$80 detected from
\ea, only three had fluences comparable {\rc to} or slightly larger than that of
{\rc the} BAT {\rc event}, while only one of the three shared a comparable
duration.

        \subsection{\large {\sc XRT Monitoring Observations}}
\swift\ observed \src\  on  15 epochs between the BAT event and
2007 January 22, for a total effective exposure time of $\sim$50\,ks (see
Table 1). For the WT mode data, the extraction region was computed
automatically by the analysis software and was a rectangle of 40 pixels
along the WT strip, centered on {\rc the} source and encompassing
$\sim 98\%$ of the Point Spread Function in this observing mode. For
PC mode observations, data were extracted from a circle with 30 pixel
radius centered on the source.  Background spectra were taken in PC
mode from annular regions (inner and outer radii of 70 and 90 pixels,
respectively), and in WT mode from a 40 pixel rectangular region in the
vicinity of the source and free of sources. The photon arrival times were
corrected to the Solar system barycenter.  {\rc Note that the absolute timing
calibration of the XRT is $\approx$200-300\,$\mu$s  (Cusumano et al. 2005).}
Spectra were extracted from the same event lists. The 1--10\,keV band was used
in the spectral fitting. Spectra were re-binned so  as to have at least 20
counts
per energy bin, so that minimum $\chi^2$ techniques could be reliably used in
the fitting.

        \subsubsection{\large {\sc Spectral analysis}}
	
We fitted the 1-10\,keV band spectral data of the first \swift\ ToO
observation of \src\ with the pre-outburst models reported in literature ({\rc
Muno et al. 2006}; Skinner et  al.  2006). The single absorbed black body (BB)
model did not  give acceptable results (reduced $\chi^2$ of 1.3 for 225 dof). A 
better fit (reduced $\chi^2 \sim$\,1.0  for 223 dof) was  obtained by using an
absorbed  BB with $kT=0.63(4)$\,keV, $R_{BB}=2.1(1)$\,km at 5\,kpc,   
plus a power-law (PL) component with photon index $\Gamma=2.3(2)$ (see Figure
\ref{specfin1}).  $N_H$ was fixed at the pre-outburst value of
1.9$\times$10$^{22}$\,cm$^{-2}$ measured with \chandra. 
Alternatively,  two BBs  with $kT_{s}=0.50(5)$\,keV
($R_{BBs}=3.2(4)$\,km) and $kT_{h}=1.1\pm_{0.1}^{0.2}$\,keV
($R_{BBh}=0.5(1)$\,km; all the uncertainties are at the 90\% level; Campana
\& Israel 2006; Israel \& Campana 2006). An F-test showed that the inclusion of
the second component is significant at {\rc the} 7$\sigma$ level. The 1--10\,keV
observed and unabsorbed fluxes were $3.7\times 10^{-11} \ergscm2$  and
$8.4\times 10^{-10} \ergscm2$, respectively, for both the two BB and the
BB+power law models (the PL component accounts for up to about 60\% of the
observed flux).  We note that the parameters of the BB component in the BB plus
PL model are significantly different from those  inferred  in quiescence
($kT_{s}=0.50$\,keV and $R_{BB}=0.36$\,km; Skinner et al. 2006; see also Section
\ref{chandrasec}).

In the following analyses we adopt the canonical BB plus PL component
to fit the entire \swift\ dataset (though there are no statistical
reasons to prefer this model over the two-BB model). We first fitted 
all together the spectra obtained in 2006, by leaving all model parameters
free except for N$_{H}$, which we kept fixed at the value determined by
\chandra\ (see above). 
Both the values of  $kT$ and $\Gamma$ were consistent with being constant in
time with marginal evidence of $\Gamma$ becoming steeper as the outburst
evolved. Therefore, we fixed both the BB temperature and the PL photon index to
the values inferred from the first \swift\ pointing (which has largest S/N) and
fitted all spectra together again.  The PL dominated the \src\ flux during the
first 10 days from the burst.  The BB component accounted for nearly 80-90\% of
the total flux during the \swift\ observations one month later; this was similar
to the spectrum in quiescence where the PL component was only marginally
detectable (see also section \ref{chandrasec}).
{\rc During the latest \swift\ observation on January 2007, the source was
caught at a 1-10\,keV observed flux level of 7.8$\times$10$^{-12}\ergscm2$ and
only the BB component was detected with a characteristic temperature of
kT=0.61(3)\,keV and radius R$_{BB}$=1.7(1) km, clearly suggesting a cooling of
the BB component as the size of the emitting region increases.}

        \subsubsection{\large {\sc Timing}} 
{\rc In this analysis we also included data from two \XMM\
and two \chandra\ observations.\footnote{For the reduction of the two \XMM\
observations we refer to Muno et al. (2006b); for the \chandra\ data reduction
and analysis (obtained in continuous clock CC33\_FAINT mode) we refer to the
procedure reported, as an example, in Rea et al. (2005).}}
We started by inferring an accurate period measurement by folding the data from
each observation at the period reported by Muno et al. (2006; see also section
\ref{chandrasec}). The majority of the resulting pulse profiles had 
three different peaks. 

The  relative phases and amplitudes were such that the signal
phase evolution could  be followed unambiguously for the observations
in the 2006 October 2-27 time interval (see Figure \ref{swiftefold}).  The
resulting phase-coherent solution had a best-fit period of P$=10.610652(1)$ s
(uncertainties here are 1$\sigma$).  TJD 13999.0 was used as reference epoch;
Details on the phase-fitting technique are given in Dall'Osso et al. (2003). A  
quadratic component was required at 4$\sigma$ confidence level in the phase
residual vs time fit. This corresponds  to $\dot{\rm P}$ = 2.4(6) $\times$
10$^{-12}$\,s s$^{-1}$. The value of $\dot{\rm P}$ is  consistent, at the
2$\sigma$ level, with the measurement of $\dot{\rm P} \simeq 1.55(50) \times
10^{-12}$\,s s$^{-1}$  reported by Woods et al. (2006), who considered the data
from 4 \chandra\ observations (from 2006 September 27 to October 28) in their
fitting, and  with the upper limit of $\sim$5.6 $\times$
10$^{-12}$\,s\,s$^{-1}$  derived by Israel et al. (2006) based on a reduced data
sample. 

Next, we compared our phase-coherent pulse profile with that derived
from observations carried out in the 2006 {\rc September} 21-26 time
interval. The signal shape showed large changes across different observations.
A one-to-one correspondence between peaks in different observations 
was found based on the fact that the reference/main peak and
the dim peak remained nearly constant, both in amplitude and relative
phase, during the whole observing period (see Figure
\ref{swiftefold}). The third peak was highly variable in amplitude
becoming comparable to, or even larger than, the reference peak after
$\sim$ 10 days from the BAT event.   The correctness of the 
reference and dim peak identification was supported by the fact
that at higher energies the shape of the post-burst pulse profile a
few days after the BAT event resembled more closely the single-peaked
pre-burst pulse profile, as shown by Muno et al.  (2006b; see also
Figure \ref{swiftefold}).  Based on the above findings we were thus
able to track the phase evolution of the reference/main  peak back to the first
\swift\ observation.
{\rc We note that techniques based on the cross-correlation of  
folded light curves from different observations proved unreliable
in our case, as they tended to favor the alignment of the highest 
peak in each observation.}

The phase-coherent solution determined for the 2006 October 2-27 time interval  
could not be extrapolated backwards to fit also the phases 
of the 2006 September 21-26 observations, not  
even by introducing higher period derivatives (up to the fourth order).   
The fit was improved by introducing an exponential term in the phase 
model, a likely signature of the occurrence of a pulsar glitch..  
The model consisting of an exponential plus a linear
and quadratic term gave  a $\chi^2 = 4.7$ for 5 dof. 
In our best fit model, the exponential component has an e-folding 
time of $\tau = 1.3(1)$~d and an amplitude of $\Delta \nu = 1.0(1) \times
10^{-5}$ Hz.  These parameters imply a large glitch, with $\Delta \nu/\nu \simeq
10^{-4}$. 

{\rc Subsequently we included three additional datasets obtained in 2007: two
short \swift\ ToO observations  carried out on January 19-22, 2007,  at the
beginning of the new visibility window of the source (see Table
2) and an archival \chandra\ DDT observations carried out on February 2, 2007. 
Unfortunately, the extrapolation of our phase coherent
solution (at 99\% confidence level) to 2007 January 19-22 observations  
resulted in a one cycle uncertainty, such that phase coherence was lost. 
The two possible phase values for the reference peak (separated by 2$\pi$)
yielded  two solutions (see Figure \ref{phaseres}). These are: (1)
P$=10.610655(1)$\,s  and  $\dot{\rm P}$ = 8.9(6) $\times$ 10$^{-13}$\,s
s$^{-1}$, and (2) P$=10.610652(1)$\,s and $\dot{\rm P}$ = 2.4(6) $\times$
10$^{-12}$\,s s$^{-1}$. 
However, the expected phase shift for the two solutions in the $\sim 2$
week-long  gap between the first \swift\ observation and the latest \chandra\
pointing is about 0.1 for solution (1) and 0.3 for solution (2). This provides
a  way to solve the ambiguity, thus recovering phase-coherence. Indeed,
our phase fitting analysis showed that the pulse phase shift accumulated is
$\sim$0.11(2), unambiguously  identifying solution (1) as the correct one.  
Fitting the whole sample of phases with a linear (P), a quadratic (\.P) and an
exponential component (glitch-like event) we derived the following final
solution:   P$=10.6106549(2)$\,s,  $\dot{\rm P}$ = 9.2(4) $\times$ 10$^{-13}$\,s
s$^{-1}$, $\tau$=1.4(1) days and an amplitude $\Delta \nu = 6.1(3) \times
10^{-6}$ Hz, corresponding to a  $\Delta \nu / \nu = 6.5(3) \times 10^{-5}$
(reduced $\chi^2$ of 1 for 14 dof). We note that the periods inferred from the
single pre- and post-burst \XMM\ observations (which have a sufficiently 
good statistics for phase fitting techniques to be used) are P$=10.61066(4)$\,s
and 10.61056(5)\,s respectively, therefore implying a
$\Delta$P$\sim$10$^{-4}$\,s. On the other hand, the  expected  $\Delta$P at the
second \XMM\ observation epoch, as inferred form the exponential decay
parameters, is $\sim$2$\times$10$^{-4}$, which is in the same range as 
the $\Delta$P reported above.}

We also tried to include the  phase point obtained from the first  \XMM\
observation, carried out $\sim$ 5 days  before the BAT event, in our solution
for the post-burst phase. We identified the single-peak of the folded lightcurve
as the reference peak we used for the phase fitting analysis (note that even a
peak misidentification in the first \XMM\ dataset would not affect the glitch
parameters or the post-glitch phase-coherent solution). 

We consider two  possibilities for the glitch phenomenology. First, 
given its large amplitude, the exponential term is assumed to amount  
to essentially the whole spin-up episode, similar to what was observed in 
the second glitch from \rxj\ (Dall'Osso et al.  2003; Kaspi et al. 2003).  
This  would require that the glitch occurred within an hour from the first
\swift\ observation. 
Alternatively, there could be a residual component of the spin-up that is
recovered over a much longer timescale (of the order
of months).  A behaviour of this type was found  in the glitch from \ea\ (Woods
et al. 2003), where the exponential recovery amounted to just $\sim$ 25\% of the
whole spin-up.  In our coherent post-glitch timing solution, a residual spin-up
component  would imply a negative linear trend in the pre-glitch phases.  The
\XMM\ data point could thus be in the position reported in Fig. \ref{phaseres} 
or shifted by multiples of $2\pi$.  

Another  possibility is that the exponential did not start
exactly at $t_0$, as it would be if the exponential had a finite rise time,
or the glitch did not occur at exactly the same time as the burst. 
We found that even a delay of 8 hours (0.33 days) would decrease the expected 
value of the residual component of the spin--up by less than a factor of 2. 
Note that the 8 hour delay corresponds to the uncertainty in the glitch epoch
determination of Woods et  al. (2003) in  the case of \ea. 

{\rc As a last step, we disregarded our identification of the peaks, based on
the
requirement that  two (out of the three) peaks remain nearly constant in
relative  phase and amplitude (see above)  and looked for further possible
timing solutions by selecting the other two peaks as the initial reference peak.
The result of this additional analysis is reported in  Figure \ref{altsol},
where the best possible alternative solution is shown. This  was obtained by
choosing the highest peak for each folded lightcurve and/or the nearest peak to
the phase extrapolation of the possible timing solution.  Fitting these phases
with a model which included a \.P component gave a reduced $\chi^2$ of 2.5
(for 11 dof) corresponding to P$=10.610654(1)$\,s and $\dot{\rm P} =  1.66(57)
\times$ 10$^{-12}$\,s s$^{-1}$ (the values are very close to those  reported by
Woods et al. 2006). The first 3-4 datapoints display a marked scattering and
give the largest contribution of the high $\chi^2$ value quoted above.
Moreover, the phase of the first \XMM\ pointing (even assuming a displacement of
2$\pi$ in the y-axis position) would be too far to be reconciled with any simple
timing solution involving only a P and \.P component. We therefore
regard this possibility as highly unlikely.}

By means of the timing solution reported above, we folded the \swift\  XRT light
curves in 13 phase-bins and inferred the corresponding root mean square (rms)
pulsed fraction for each of them. In Figure \ref{rmsok} (lower panels)  we
report both the rms pulsed fraction as well as the pulsed intensity (in counts
per second) as a function of the average source intensity.  The 
fractional rms rose towards the pre-outburst value of
$0.54$ toward the end of the monitoring observations. However, the countrate of 
the pulsed component remained nearly constant throughout the outburst,
despite a factor of $\approx$3 variation in the average flux. 

A search both in XRT and BAT monitoring data for additional X--ray bursts like
the one detected by BAT gave negative results. 

        \subsection{\large {\sc \chandra\ archival datasets}}
	\label{chandrasec}
	
We retrieved  from the \chandra\ archive two observations of Westerlund 1
carried out with the ACIS-S imaging array on 22-23 May 2005 and 18-19 June
2005  with effective exposure time  of 18.8\,ks and 38.5\,ks, respectively. The
observations were obtained in faint-event mode using a 3.2 s frame time.
Details on the reduction were reported in Muno et al. (2006) and Skinner et al.
(2006). In our re-analysis we focused on a pulse phase-resolved spectroscopy
study of the longest observation. 

Source photon arrival times were extracted from circular regions with a radius
of 2\arcsec, which included about 95\% of the source photons, and were
corrected to the barycenter of the solar system.  The best period of the longest
observation was determined to be  10.61068(11)\,s, by fitting the phases of the
modulation  obtained over 4  consecutive intervals each $\sim 10^{4}$~s long
duration (uncertainties are at the 90\% confidence level). 
The value was consistent with the period reported by Muno et al. (2006).

By folding the lightcurve at the best period reported above, a nearly
sinusoidal shape of the modulation was found with a large pulsed fraction
($\sim$78\%$\pm$4\%; semi amplitude of the modulation divided by the mean 
source countrate; 52\% in term of fractional RMS). Moreover, the pulsed
fraction was consistent with being slightly  energy--dependent in the  soft and
hard bands (pulsed fractions of 70\%$\pm$5\% 88\%$\pm$5\% for the 0.5-2 and
2-10 keV bands, respectively; 50\%$\pm$7\% and 59\%$\pm$7\% in term of 
fractional RMS).  There {\rc were} also indications that the minimum of the
modulation was  shallower at high energies (see Figure
\ref{efold}).  Then, we identified 4 phase intervals (0.0-0.2, 0.2-0.6, 0.6-0.8
and 0.8-1.0; 
see Figures \ref{efold} and \ref{pps}) within which a single spectrum was
obtained.  

We fitted the 1-10\,keV phase-averaged spectrum with the BB and BB plus PL
models.  The inclusion of the PL component was found to be significant
at the 98\% confidence level. The best fit (reduced $\chi^2\sim 1.14$
for 43 dof) was obtained for $\Gamma = 3.5\pm^{1.3}_{0.3}$,
$kT\sim0.49\pm0.1$\,keV and $R_{BB}\sim0.4$\,km. The PL component
accounted for 60\% of the total flux.  These results are similar to
those reported by Skinner et al. (2006). {\rc It should be noted,
however, that a BB spectrum might not necessarily be the most
appropriate description of the spectrum. Indeed, magnetized, light
element atmospheres are often used to fit the spectra of magnetic
neutron stars. Furthermore, general relativistic corrections (which
are not accounted for in the BB model) also play a role in 
the observed spectrum.  In the case of \src, fits using a magnetized
atmosphere and the appropriate general relativistic corrections
were reported by Skinner et al. (2006). These yielded a slightly
lower effective temperature and a slightly larger radius
of the emitting region. However, the statistics were not sufficient to
favor this model over the simple BB.}

Finally, we  carried out phase-resolved spectroscopic by keeping 
$N_{H}$ and $\Gamma$ fixed at the values inferred in the phase-averaged
spectrum. The results of the spectral fitting are reported in Table 2
and shown in Figure \ref{pps} (lower panel). An F-test gave a probability of
99.2\%  for the  PL component  to be significant in  the four phase interval
spectra together.  Based on the above analyses, we note that the presence of the
PL component was corroborated by the fact that the countrate ratio between the
pulse minimum and maximum were approximatively equal to the  BB/PL flux ratios
inferred from the PPS analysis  (see first two panels of Figure \ref{pps}). To
further test this hypothesis we folded the lightcurve considering only photons
at energies above 4.5\,keV, where  the contribution from the BB component 
was negligible at pulse minimum. The shape of the modulation changed drastically
with respect to that at lower energies, showing an asymmetric profile  with a
flat  0.2 phase-long  minimum consistent with zero countrate (see Figure
\ref{efold}). 

We note that the unusually small size inferred for the BB emitting region
(radius of 270\,m) {\rc is consistent with the large pulsed flux.  
The BB component is emitted from a hot spot on the NS surface, and 
the pulsed fraction reaches its minimum as the BB
emitting region gets completely occulted for a portion of the star rotation
period. Therefore, at pulse minimum, the spectrum is dominated by the PL
component.}

	  \section{\Large {\sc Discussion}}
In the following we compare the \swift\ results of \src\ with those of other
AXPs that showed a similar behavior. 
	  
A large glitch ($\Delta \nu / \nu \sim 4 \times 10^{-6}$) was detected from \ea,
nearly simultaneously with several intense (total fluence of $\sim 3 \times
10^{-8}$\,erg\,cm$^{-2}$) and short SGR-like bursts (Woods et al. 2004, Kaspi et
al. 2003). Our \swift\ and \XMM\ coherent timing solution of \src\  implies that
an even larger glitch ($\Delta \nu / \nu = 6.5 \times 10^{-5}$) occurred within
0.8 days from the burst epoch.  The glitch was more than one order of magnitude
larger than  previous glitches detected from AXPs, or any other type of pulsar.
This holds true even considering the minimum value of  the residual component of
the spin--up for a possible long-term glitch recovery component.
The glitch effects ($\Delta \nu / \nu \approx  6 \times 10^{-5}$) were 
recovered  for the most part  over less than 2 days, although the exponential
component was still  detectable after one week.  In this respect the exponential
decay time and its amplitude are, taken separately, within the wide range found
for radio pulsars, but they were never observed to occur together (Wong et
al. 2001; Hobbs et al. 2002).  On the other hand, the combination of quick
recovery and large amplitude is similar to  that  observed in other AXPs/SGRs
(Dall'Osso et al. 2003; Woods et al. 2004).   We also detected a secular
spin-down $\dot{\rm P}$ = 9.2(4) $\times$ 10$^{-13}$\,s s$^{-1}$, implying a
dipole field strength of $\simeq  1 \times 10^{14}$ Gauss (assuming a neutron
star radius of 10\,km and a mass of 1.4$_{\odot}$), a value not dissimilar from
that inferred for other AXPs. The long-term recovery of the glitch can be
reasonably expected to be characterized by a second  derivative of the period,
which might be revealed by future monitoring  observations of the source (as in
the case of \rxj where it was revealed  over a $\sim$500 day timescale). The
detection of such a component  would provide a further confirmation of the
presence and amplitude of the  residual component, especially once the recovery
timescale is known (cf.  Dall'Osso et al. 2003, $\S$ 6.3 and references
therein).
{\rc From the energetic point of view we note that whatever was  the cause of 
the burst, the glitch  and the outburst, it released 
approximately 5 $\times$ 10$^{41}$\,ergs during the first 130 days: 
 0.004\% of which  has been emitted during the burst, 6\% has been stored in
the star during the glitch and lost again (either through internal dissipation
or emission) within a couple of days.  More than 90\% of the total went to
increase the persistent (mostly unpulsed) emission. Future monitoring
observations of the source flux decay until the quiescence level will allow us
to quantify the total energy budget of the outburst.}

The prompt X-ray afterburst properties of \src\ also are amazingly
similar to those of \ea. In fact, in the case of \ea, a drop
in the pulsed fraction was observed together with a change in the
pulse shape (from 25\% to 15\% rms) quickly recovering (within a week)
towards the pre-burst shape and pulsed fraction. Similarly, the
\src\ pulse shape varied from a nearly sinusoidal shape (before the
burst) to a double peaked one (after 1 day) and to a triple-peak shape
(for epochs later than 2 days), while the pulsed fraction dropped from
a value of $\sim80\%$ (as recorded by an \XMM\ observation few days
before the burst) to $\sim10\%$ few hours after the BAT event. Since
then the pulsed fraction has been recovering  towards its pre-burst value.
Moreover, the flux decay of \src\ is reminiscent of that of \ea,
e.g. a power-law $F \propto t^{\alpha}$, with ${\alpha}$ index of
--0.28$\pm$0.05 after the first day from the BAT event (compared with
the --0.22 in \ea).  It is also apparent that the PL component decayed
more rapidly (${\alpha}$ index of --0.38$\pm$0.11; 90\% uncertainty)
than the BB flux (${\alpha}$ index of --0.14$\pm$0.10).  This implies 
that the cooling timescale of the hot spots on the neutron star's surface
is longer than that of the region responsible for the power law, 
likely an active coronal region (used to account for the broadband
non-thermal PL component; Beloborodov \& Thompson 2006).  Regardless
of the scenario, the pulsed emission was mainly accounted for by the BB,
and the increase of the pulsed fraction in the post-burst phases was
caused by the increase in the fractional contribution of the BB flux as the
nearly-unpulsed PL component decayed.  Further evidence in favor of this comes
from the rise in  the fractional RMS as a function of time. This was well-fitted
by a power-law with index ${\alpha}$ of +0.38$\pm$0.11, which is opposite to the
slope of the decay of the  amplitude of the power-law component.

The BB component also appears to play an important role in the quiescent state
of this AXP, as shown by  our PPS study in Section \ref{chandrasec}, where the 
timing and spectral properties were found consistent with the  100\% pulsed
component being totally accounted for by the BB component. This very high degree
of modulation is consistent with the  small size of the region  for the BB
component, if this originates at the NS surface and  is periodically
self-eclipsed by rotation. For a typical star of radius $\sim 10$ km, and a
distance of $\sim 5$\,kpc, the angular size BB emitting spot is only a few
degrees. Although light-bending effects contribute to substantially suppress the
degree of modulation, for a spot of such a small extent there exist 
geometries for which complete occultation can be achieved for a part of the star
rotation (1983; see also Riffert \& Meszaros 1988 and \"Ozel 2002; DeDeo et al.
2001). 

It is interesting to note that, though formally not required from a
statistically point of view, there were indications of the presence of  a soft
excess  in the  \chandra\ spectrum corresponding to  the pulse minimum (where
the BB from the hot spot is absent; see central panel of Figure \ref{pps}). By
modeling this excess with a BB component we obtained an upper limit of
$kT \leq$0.15keV  and radius consistent with the NS size.  If confirmed by
future observations of \src\ in quiescence, this latter component might 
provide evidence for the thermal radiation emitted from the whole  NS surface 
(similar to what we observed in the quiescent spectrum of \xte).

One of the main differences (in the intial phases) between the outbursts from  \src\
and \ea\
is that the quiescent flux  component was a factor of 100 larger in
\ea, so that the  relative amplitude of the outburst  was smaller. 
In fact, the luminosity of these two sources one day after their 
respective glitches were the same to within a factor of 2 
(\ea\ was slightly brighter).  In this respect,  the outburst properties of the
two AXPs are more similar than they would appear at first sight. Indeed, the
longer timescales over which the pulsed component rms recovers its initial value
 in \src\ is likely due to   the fainter quiescent level of this source with
respect to that in \ea\ (where the rms is completely recovered, i.e. it reaches
the pre-outburst  value, in less than a week). 

Particularly relevant is the flare event detected in 2003 from \ee, another
historical AXP. A clear anti-correlation between pulsed
fraction and intensity of the source was detected and found to be
accompanied by relatively small variations in the phase-averaged
spectrum (Tiengo et al. 2005). The presence of a hard component was
detected in the PPS study at the maximum of the pulses. A monitoring
campaign of \ee\ carried out by RXTE showed that this phenomenology
was also accompanied by a large outburst of the source flux above
its previous average value, which lasted for about 2 years. However,
no burst was detected close to the outburst onset epoch (Kaspi et
al. 2005). In this case the timescale over which the pulsed fraction
was recovered was of the order of the outburst duration itself instead
of one week (as observed in \ea).  This event has several similarities
with the one currently observed in \src.

Finally, it is likely premature to
draw possible similarities between the outbursting behavior of \src\
during the first month and that of \xte, (Ibrahim et al.  2004;
Gotthelf et al. 2004; Israel et al. 2004; Rea et al. 2004). Both the
timing and spectral properties appear to be different: the pulsed fraction
of \xte\ increased its value from a pre-outburst upper limit of 20\%
(in the 0.1-2.5\,keV band) to a 1-10\,keV measurement of $\sim$53\%
after about one year, and  decreased since then.  Moreover, the quiescent
spectrum differed, in that it was dominated by a
very soft and extended BB in \xte, and by a relatively hard and
especially small radius BB in \src. These differences may result from
different viewing angles of and/or different burst emitting regions.   

	  \section{\Large {\sc Conclusions}}
On September 21, 2006, the candidate AXP \src\ emitted a short and rather
intense burst that was promptly detected by the \swift\ BAT.   Together with the
burst,  large changes in the timing and spectral properties of the persistent
component were detected and seen evolving during the subsequent weeks. In
particular, the \swift\ XRT monitoring (plus two proprietary \XMM\  and two 
archival \chandra\ observations)  allowed us to find the following: 

\begin{itemize} 

\item  The pulse phase evolution  is consistent with the occurrence of a large
glitch ($\Delta \nu / \nu \sim 10^{-4}$),  the largest ever detected from a
neutron star. The glitch was recovered over a timescale of 1.4\,d, though its
effects were present in the pulse phases until approximately one week after the
glitch epoch. We also detected a quadratic component in the pulse phases 
corresponding to a $\dot{\rm P}=9.2(4) \times 10^{-13}$ s s$^{-1}$. 

\item The first 1-10\,keV \swift\ XRT spectrum was measured $\sim$13 hours
after the burst detection and  showed, in addition to a $kT \sim0.65$\,keV
blackbody ($R_{BB}\sim1.5$\,km), a $\Gamma\sim2.3$ power-law component
accounting for about 50\% of the  observed flux. 

\item The flux decay of \src\ is well described by the function $F
\propto t^{\alpha}$, with index ${\alpha}$ of --0.28$\pm$0.05 (similar
to the case of the 2002 \ea\ burst-active phase). Moreover, we
found that the PL component decays more rapidly (index ${\alpha}$ of
--0.38$\pm$0.11; 90\% uncertainty) than the BB flux (index ${\alpha}$ 
of --0.14$\pm$0.10).

\item The pulsed fraction of the 10.61\,s pulsations was seen to drop from a
value of $\sim80\%$ (as recorded by an \XMM\ observation a few days before the
burst) to $\sim10\%$ a few hours after the BAT event. The spectral and
timing analysis clearly show that only the blackbody component is responsible
for the pulsed flux (at least during the initial phases of the outburst). 

\item Archival \chandra\ data analysis revealed that the modulation in
quiescence is 100\% pulsed at energies above $\sim$4\,keV and consistent with
the (unusually small-size) blackbody component being occulted by the neutron
star as it rotates.

\item A comparison of the  properties of \src\ with those of other AXPs which
showed similar behaviour  confirmed that outbursting events of this  kind are
more common  than previously thought.
\end{itemize}	  

All these results confirmed unambiguously that \src\ is a transient and bursting
AXP, showing an unusually high pulsed fraction level in quiescence.  Studying
how the source will return from its post-burst/glitch  timing and spectral
properties to the quiescent ones  might help revealing  the mechanisms behind
the outbursts of AXPs.  Finally, the  BAT detection of bursts from \src\ opens a
new perspective for detecting  bursts from known AXPs, and for identifying new
AXPs/SGRs  with \swift.
	  
	\acknowledgments

This work is partially supported at OAR through Agenzia Spaziale Italiana (ASI),
Ministero  dell'Istru\-zione, Universit\`a e Ricerca Scientifica e Tecnologica
(MIUR -- COFIN), and Istituto Nazionale di Astrofisica (INAF) grants. We
acknowledge financial contribution from contract ASI-INAF I/023/05/0. We thank 
Neil Gehrels for approving the \swift\ ToO observation program and Norbert
Schartel for  approving the \XMM\ post-burst observation through the
Director's Discretionary Time program. We thank Patrizia Romano for her help in
the quick look of the TDRSS data of the BAT event. We also thank Nanda Rea,
Andrea Possenti, Marta Burgay, Diego G\"otz and Peter Woods for useful
discussions.

{\it Facilities:} \facility{Swift (BAT)}.
{\it Facilities:} \facility{Swift (XRT)}.
{\it Facilities:} \facility{Chandra (ACIS-S)}.
{\it Facilities:} \facility{XMM-Newton (EPN)}.

\clearpage


\begin{figure}
\includegraphics[angle=-90,scale=.90]{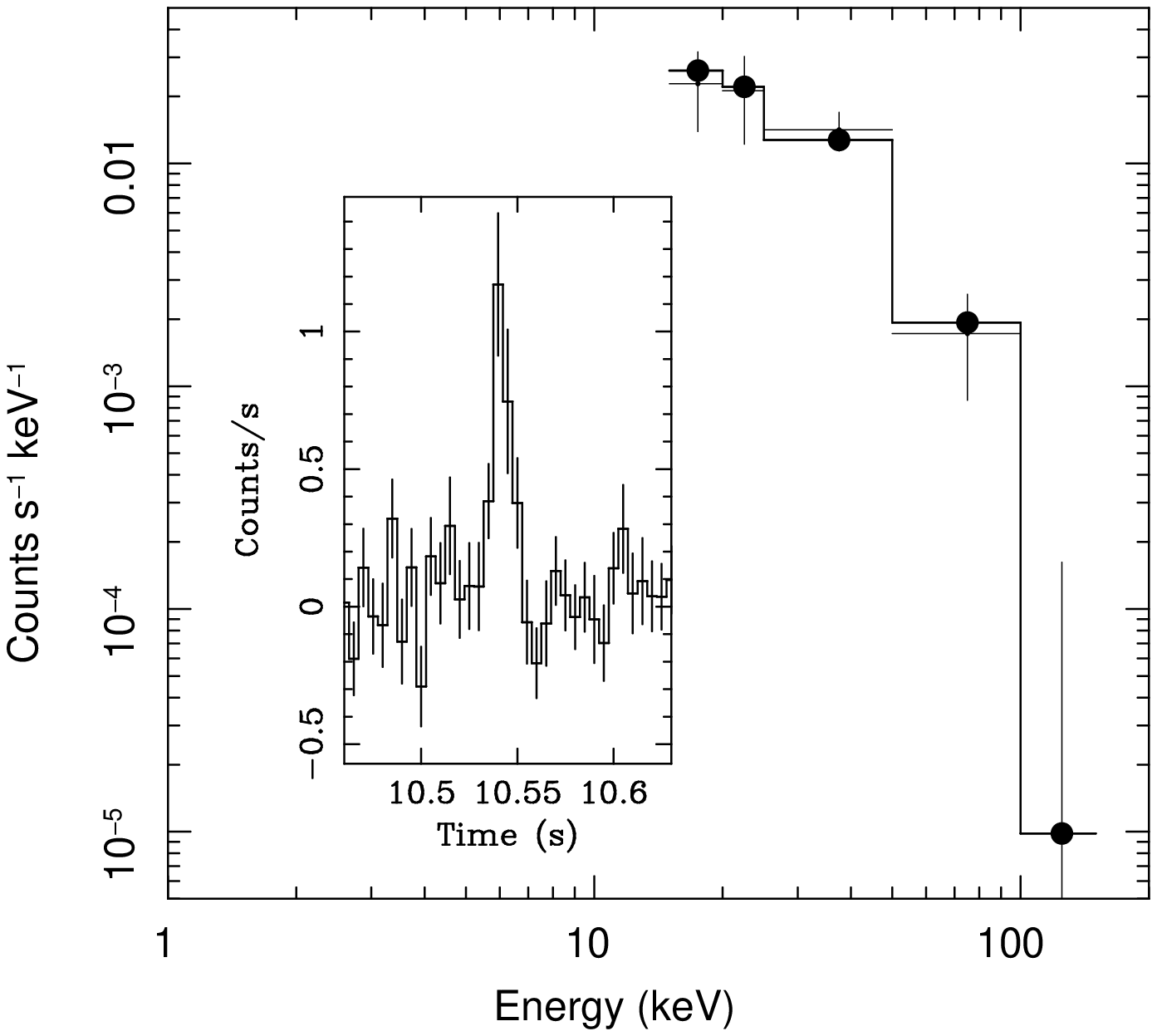}
\caption{The 15-150\,keV \swift\ BAT spectrum of the 20ms long burst detected
from \src\ on 21st September 2006, together with the 5ms-binned BAT lightcurve
in the time interval around the trigger (inset).\label{BATevent}}
\end{figure}
\begin{figure}
\includegraphics[angle=-90,scale=0.9]{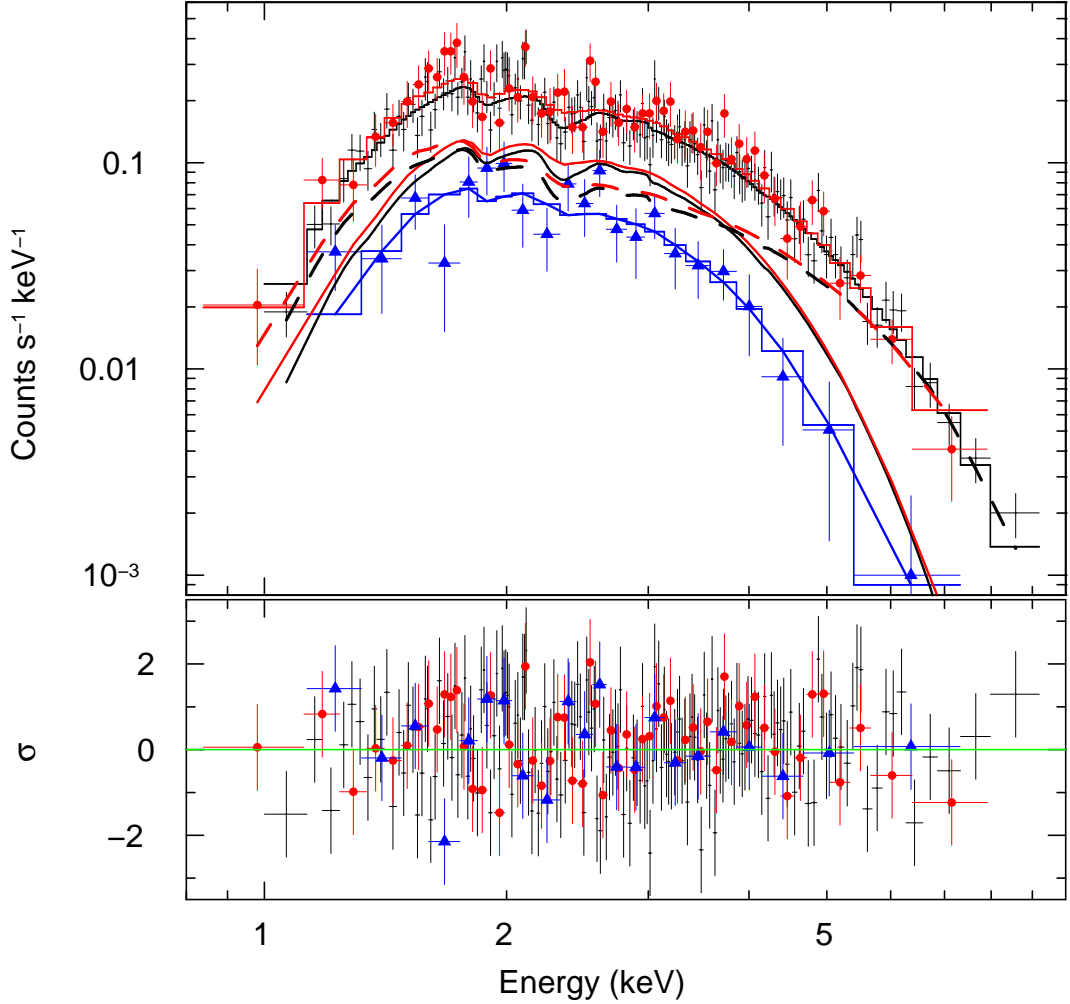}
\caption{The 1-10\,keV \swift\ XRT spectra obtained during the first  ToO
observation started 13 hours after the BAT event (2006 September 21; ), and the
most recent one started 120 days later (2007 January 19-22; filled triangles)
are shown together with the model residuals. Both the WT (filled circles) and PC
mode spectra are displayed for the first observation and fitted with the BB plus
PL (marked with solid  and stepped lines, respectively) model discussed in the
text. During the January 2007 observation (only WT mode data were obtained) the
PL component was not present anymore.
\label{specfin1}}
\end{figure}
\begin{figure}
\includegraphics[angle=-90,scale=.60]{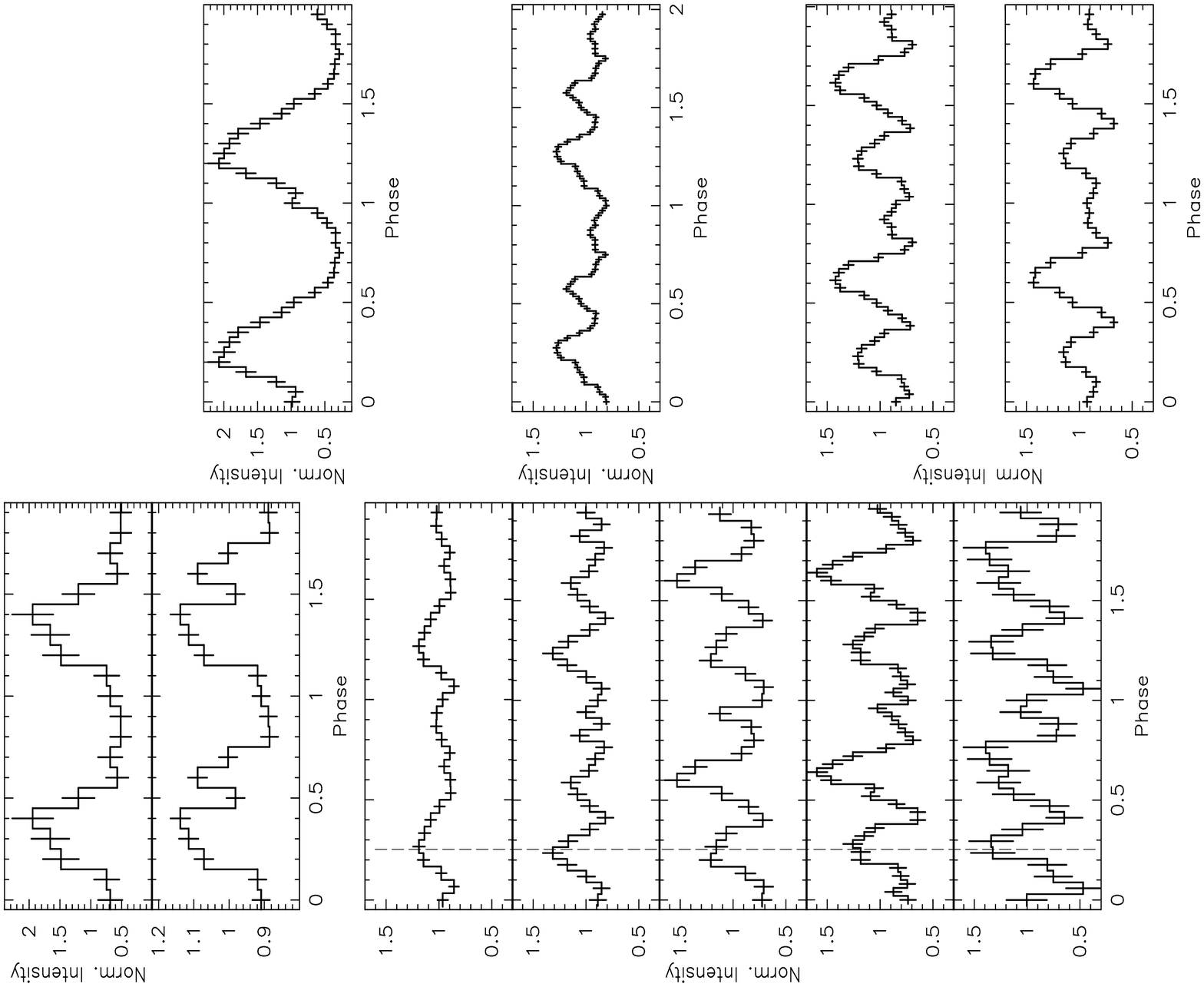}
\caption{3.5-10\,keV \XMM\ PN lightcurves (upper left panels; 5 days before and
2 days after the BAT event), the 1-10\,keV 
\swift\ XRT  lightcurves referring to $t_0+0.7$d, $t_0+1.8$d,  $t_0+12$d,
$t_0+[15;37]$d and $t_0+[120]$d from top to  bottom (lower left panels). In the
right panels are shown the 1-10\,keV \XMM\ PN and \chandra\ ACIS-S lightcurves
referring to $t_0-5$d,  $t_0+1.7$d,   $t_0+38$d and $t_0+135$d
folded by using the $P-\dot{\rm P}$ coherent timing solution discussed in the
text . The peak at phase $\sim0.25$ marked by the stepped line represents our
reference peak. The peak at phase $\sim 0.6$ is the highly variable component we
identify.\label{swiftefold}}
\end{figure}

\begin{figure}
\includegraphics[angle=-90,scale=.80]{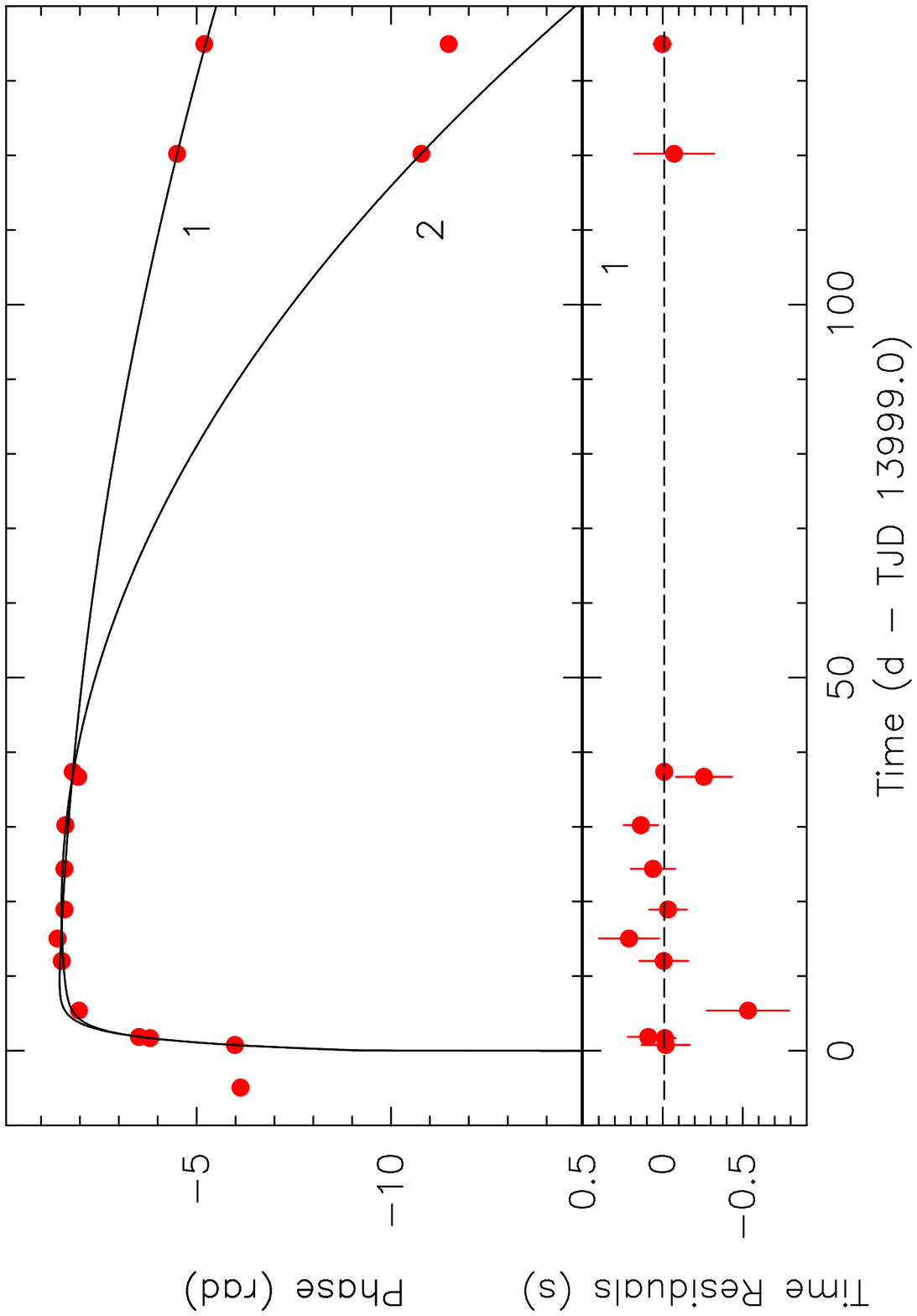}
\caption{Phases (upper panel) of the \swift\ XRT,  \XMM\ and \chandra\
observations of \src\ (see the text for the discussion concerning the \XMM\
pre-burst phase position): a large and quick decaying component is clearly
present. Time residuals (lower panels) in seconds of the above datapoints with
respect to the  phase coherent P-$\dot{\rm P}$ timing solution discussed in the
text and 
including an exponential component. Note that the first \XMM\ point (at
day -5)  would be at the reported phase only in the hypothesis that the pre- and
post-glitch parameters are similar (see the text for discussion).
\label{phaseres}}
\end{figure}
\begin{figure}
\includegraphics[angle=-90,scale=.80]{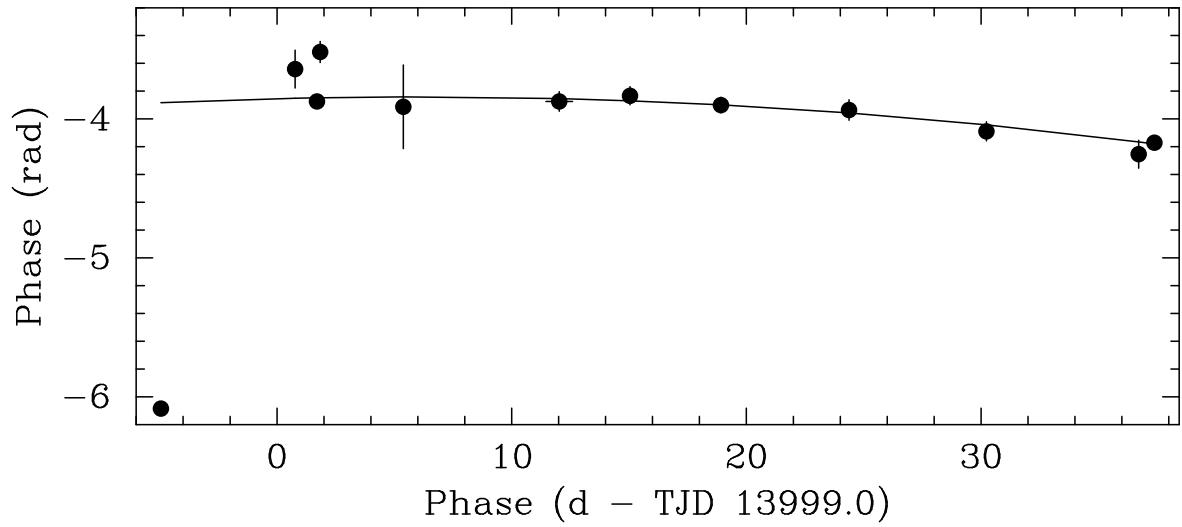}
\caption{The best alternative solution we found by removing our initial
hypothesis of having two peaks nearly constant through all the observations (see
the text for details). Superimposed is the fitted model including a linear and
quadratic term (without considering the pre-burst \XMM\ observation). 
\label{altsol}}
\end{figure}
\begin{figure}
\includegraphics[angle=-90,scale=.80]{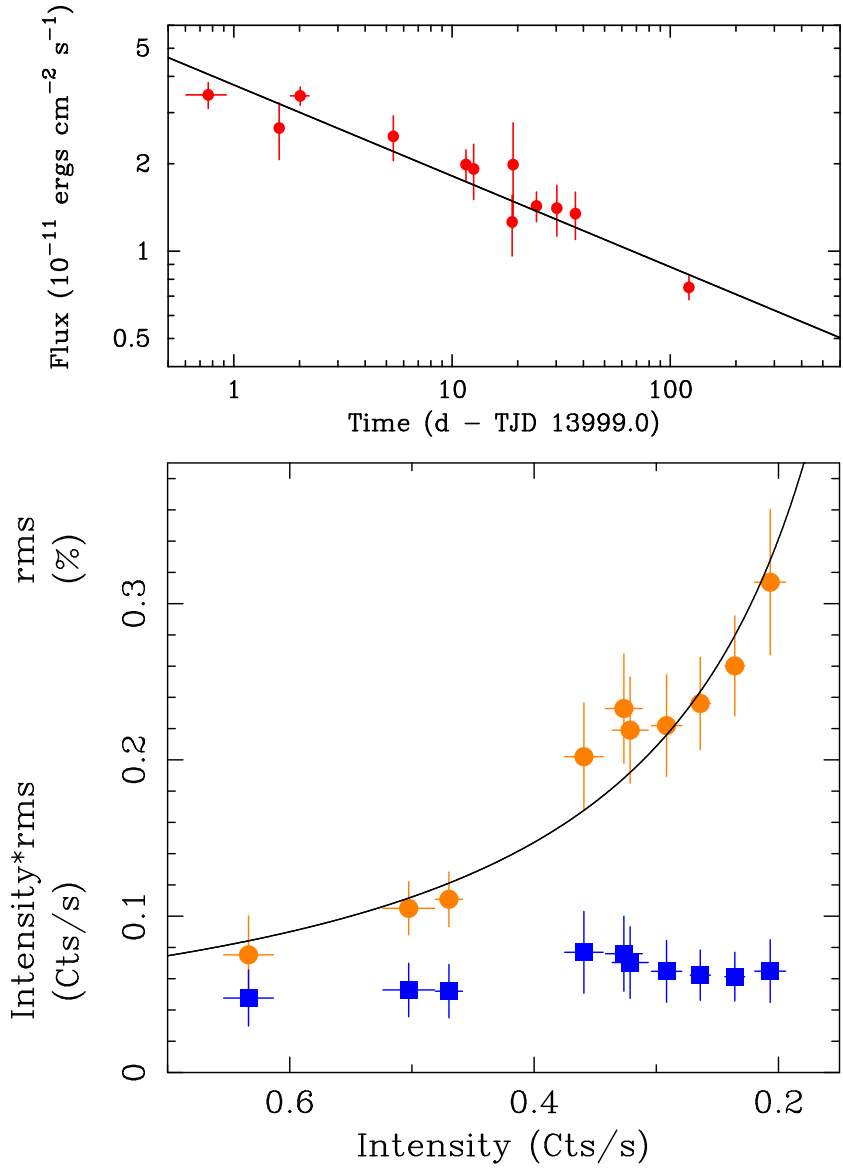}
\caption{The 1-10\,keV \swift\ XRT flux lightcurve of \src\ with the
power-law component superimposed used to model the decay 
(upper panel; see text for details).
The 10.6\,s signal fractional RMS (filled circles) and pulsed intensity (filled
squares) as a function of the average source countrate (lower panel).
\label{rmsok}}
\end{figure}

\begin{figure}
\includegraphics[angle=-90,scale=1.10]{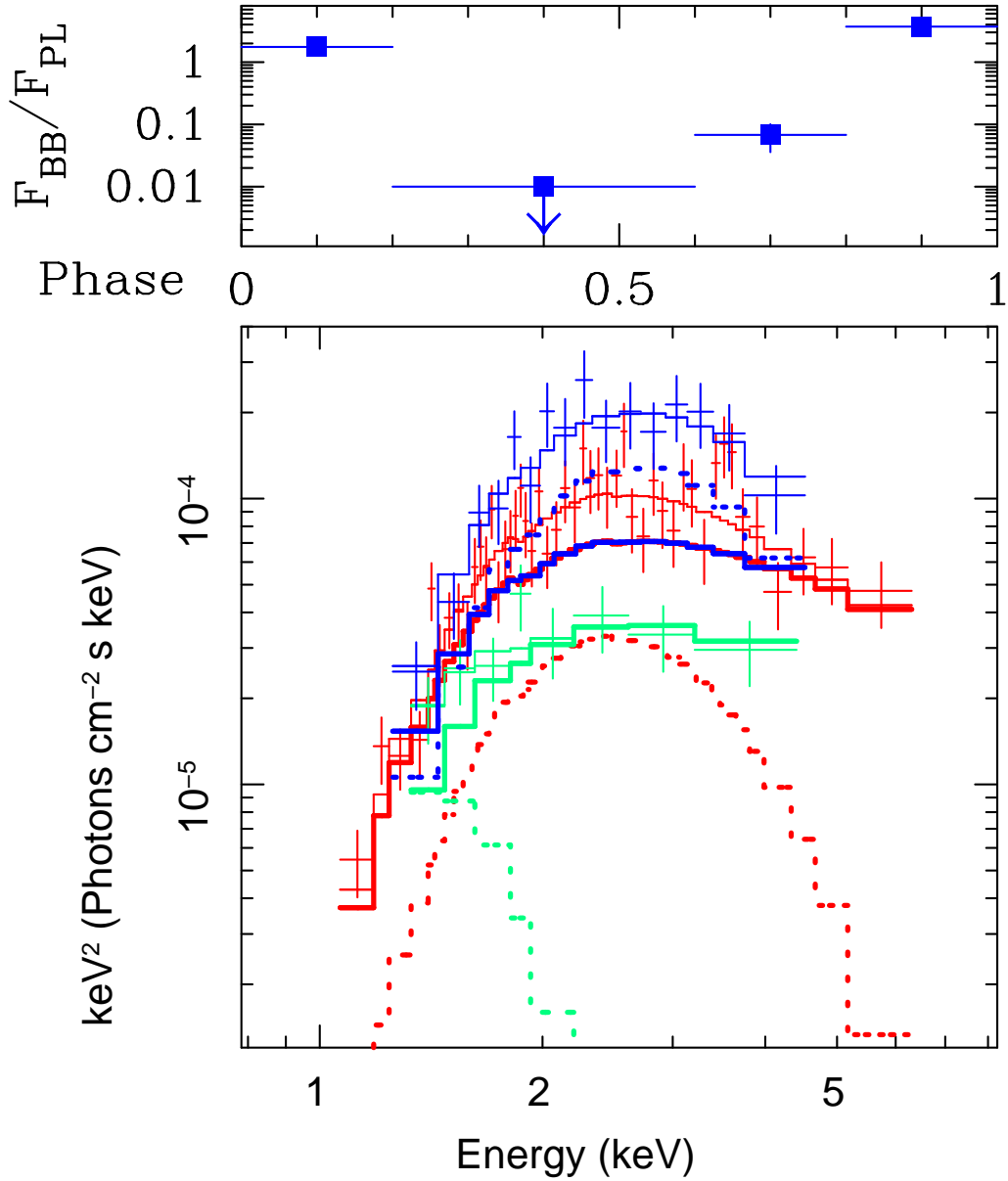}
\caption{ACIS-S \chandra\ Pulse Phase Spectroscopy results (only 3 phase
intervals are reported for clarity) obtained  from archival \chandra\ data of
\src\ by assuming a BB plus PL spectral model. The BB components are marked by
the dotted lines, PL with the solid ones. The flux ratio of the two components
is
also shown in the upper panel as a function of pulse phase.\label{pps}}
\end{figure}

\begin{figure}
\includegraphics[angle=-90,scale=1.10]{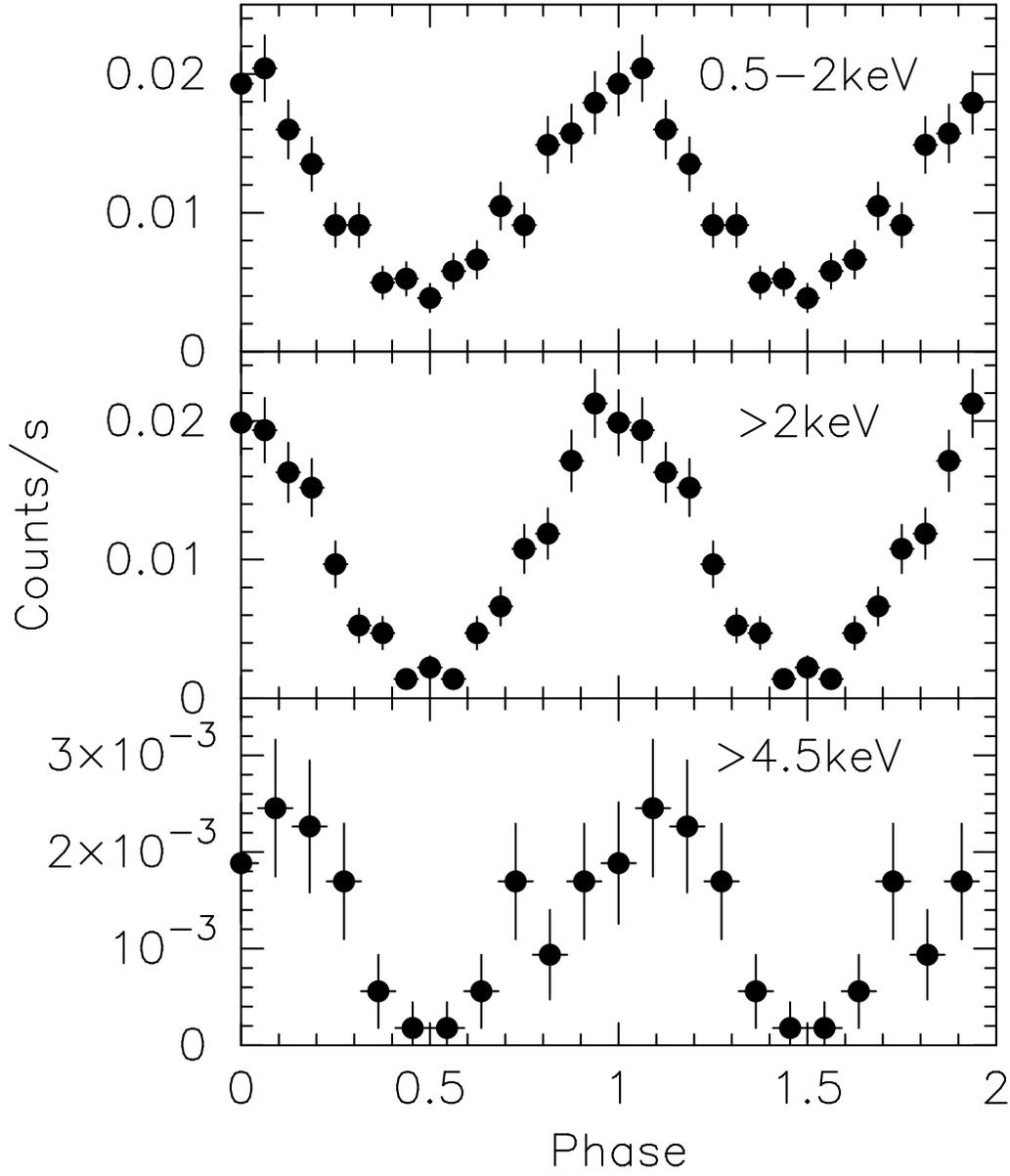}
\caption{0.5-10\,keV \chandra\ ACIS-S folded lightcurves at different energies
of \src. Phase 0 is arbitrarily set to TJD 13539.0  .\label{efold}}
\end{figure}


\clearpage

\begin{deluxetable}{lllll}

  \tablewidth{0pc}              
  \tabletypesize{\scriptsize} 
  \tablecaption{\label{alldata}\swift\ Observation Log for \src.} 
  \tablehead{
\colhead{Sequence ID} &    \colhead{Obs/Mode} &    \colhead{Start time  (TDB)} &
  
\colhead{End time (TDB)} & \colhead{Exposure}  \\
\colhead{} &  \colhead{} &    \colhead{(yyyy-mm-dd hh:mm:ss)} &
\colhead{(yyyy-mm-dd hh:mm:ss)} &    \colhead{(s)} 
}
 \startdata     
00230341000\tablenotemark{a} &       BAT/EVENT &        2006-09-21 01:34:11    &
2006-09-21T01:34:54     &       43      \\

\hline
00030806001 & XRT/WT & 2006-09-21 14:29:13 & 2006-09-21 22:10:33 &
1919.9   \\
00030806001 & XRT/PC & 2006-09-21 14:29:29 & 2006-09-21 22:27:44 &
7736.6   \\
00030806002 & XRT/WT & 2006-09-22 14:38:22 & 2006-09-22 14:51:10 &
766.9    \\
00030806003 & XRT/WT & 2006-09-22 19:39:09 & 2006-09-23 05:04:12 &
4910.0   \\
00030806003 & XRT/PC & 2006-09-22 19:39:17 & 2006-09-23 01:44:13 &
1765.6   \\
00030806004 & XRT/WT & 2006-09-26 06:36:51 & 2006-09-26 11:35:03 &
1250.0   \\
00030806004 & XRT/PC & 2006-09-26 08:12:58 & 2006-09-26 11:42:46 &
2482.3   \\
00030806006 & XRT/WT & 2006-10-02 11:04:54 & 2006-10-02 17:09:01 &
1977.6   \\
00030806007 & XRT/WT & 2006-10-03 12:19:42 & 2006-10-03 14:12:54 &
2034.1   \\
00030806008 & XRT/WT & 2006-10-05 23:59:24 & 2006-10-06 01:48:39 &
2158.9   \\
00030806009 & XRT/WT & 2006-10-09 17:51:53 & 2006-10-09 22:51:11 &
3521.8   \\
00030806010 & XRT/WT & 2006-10-10 00:09:50 & 2006-10-10 02:13:09 &
2829.4   \\
00030806011 & XRT/WT & 2006-10-15 05:38:04 & 2006-10-15 12:21:36 &
5617.9   \\
00030806012 & XRT/WT & 2006-10-21 02:56:44 & 2006-10-21 08:07:58 &
5508.0   \\
00030806013 & XRT/WT & 2006-10-27 16:10:55 & 2006-10-27 19:42:44 &
2816.6  \\
 00030806014 & XRT/WT & 2007-01-19 04:05:59 & 2007-01-19  06:00:28&
2051.8    \\
00030806015 & XRT/WT & 2007-01-22  01:07:57& 2007-01-22 04:44:42 &
3817.7  
    \enddata 
    \tablenotetext{a}{BAT trigger.} 
   \end{deluxetable}  

\clearpage 

\begin{table}
\begin{center}
\label{ppspar}
 
\caption{ PPS analysis of \chandra\ data for \src.}
\vspace{4mm}
\begin{tabular}{lcccc}
 \tableline\tableline
Spectral & \multicolumn{4}{c}{Pulse Phase Interval}\\
Parameter              &  A            &   B (minima)  &   C   &  D (maxima) 
\\\tableline
N$_H$ (cm$^{-2}$)& \multicolumn{4}{c}{Fixed at phase-averaged value of 1.9} \\
$kT_{BB}$ (keV)  & 0.50$\pm$0.02 & $<$0.15 & 0.65$\pm$0.15& 0.54$\pm$0.02\\
R$_{BB}$  (km @ 5kpc)  & 0.4$\pm0.1$& $<$20 & 0.2$\pm$0.1 & 0.44$\pm$0.06\\
PL $\Gamma$ & \multicolumn{4}{c}{Fixed at phase-averaged value of 3.5} \\
$F_{BB}$ ($10^{-13}\ergscm2$) & 1.18$\pm$0.18& $<$0.07& 0.17$\pm$0.03&
2.41$\pm$0.36\\
$\chi^2_{\nu}$ & \multicolumn{4}{c}{0.9 by fitting the 4 spectra together} \\
\tableline
\end{tabular}
\end{center}
\end{table}
%


\begin{thebibliography}{}

\bibitem[Beloborodov (2006)]{Andrei 2006} Beloborodov, A.M.  \& Thompson, C.
2006, \apj, in press (astro-ph/0602417)

\bibitem[Burgay et al.(2006)]{2006ATel..903....1B} Burgay, M., Rea, N., 
Israel, G.L., \& Possenti, A.\ 2006, The Astronomer's Telegram, 903


\bibitem[Camilo et al.(2006)]{2006Natur.442..892C} Camilo, F., Ransom, 
S.~M., Halpern, J.~P., Reynolds, J., Helfand, D.~J., Zimmerman, N., \& 
Sarkissian, J.\ 2006, \nat, 442, 892 

\bibitem[Campana \& Israel(2006)]{2006ATel..893....1C} Campana, S., \& 
Israel, G.~L.\ 2006, The Astronomer's Telegram, 893

\bibitem[Cusumano et al.(2005)]{2005SPIE.5898..377C} {\rc Cusumano, G., et al.\ 
2005, \procspie, 5898, 377 }

\bibitem[Dall'Osso et al. (2003)]{2003ApJ...599..485D} Dall'Osso, S., 
Israel, G.~L., Stella, L., Possenti, A., \& Perozzi, E.\ 2003, \apj, 599, 
485 

\bibitem[DeDeo et al.(2001)]{2001ApJ...559..346D} DeDeo, S., Psaltis, D., 
\& Narayan, R.\ 2001, \apj, 559, 346 

\bibitem[Duncan \& Thompson(1992)]{1992ApJ...392L...9D} Duncan, R.~C., \& 
Thompson, C.\ 1992, \apjl, 392, L9 

\bibitem[Gavriil et al.(2002)]{2002Natur.419..142G} Gavriil, F.~P., Kaspi, 
V.~M., \& Woods, P.~M.\ 2002, \nat, 419, 142 	

\bibitem[Gavriil et al.(2004)]{2004ApJ...607..959G} Gavriil, F.~P., Kaspi, 
V.~M., \& Woods, P.~M.\ 2004, \apj, 607, 959 

\bibitem[Gotthelf et al.(2004)]{2004ApJ...605..368G} Gotthelf, E.~V., 
Halpern, J.~P., Buxton, M., \& Bailyn, C.\ 2004, \apj, 605, 368 

\bibitem[Gotthelf \& Halpern(2005)]{2005ApJ...632.1075G} Gotthelf, E.~V., 
\& Halpern, J.~P.\ 2005, \apj, 632, 1075 

\bibitem[Gotz et al.(2006)]{2006ATel..953....1G} {\rc Gotz, D., et al.\ 2006, 
The Astronomer's Telegram, 953}

\bibitem[Hill et al.(2004)]{xrtmodes} 
	Hill, J.~E.,  Burrows, D.\ N., Nousek, J.\ A., et al.\ 2004, \procspie, 5165,
217 

\bibitem[Hobbs et al.(2002)]{2002MNRAS.333L...7H} Hobbs, G., et al.\ 2002, 
\mnras, 333, L7 

\bibitem[Ibrahim et al.(2004)]{2004ApJ...609L..21I} Ibrahim, A.~I., et al.\ 
2004, \apjl, 609, L21 

\bibitem[Israel \& Stella(1996)]{1996ApJ...468..369I} Israel, G.~L., \& 
Stella, L.\ 1996, \apj, 468, 369 

\bibitem[Israel et al.(2004)]{2004ApJ...603L..97I} Israel, G.~L., et al.\ 
2004, \apjl, 603, L97 

\bibitem[Israel \& Campana(2006)]{2006ATel..896....1I} Israel, G.~L., \& 
Campana, S.\ 2006, The Astronomer's Telegram, 896

\bibitem[Israel et al. (2006)]{2006ATel..932} Israel, G.~L., 
Dall'Osso, S., Campana, S., Muno, M., \& Stella, L.\ 2006, The Astronomer's
Telegram, 932

\bibitem[Kaspi et al.(2000)]{2000ApJ...537L..31K} Kaspi, V.~M., Lackey, 
J.~R., \& Chakrabarty, D.\ 2000, \apjl, 537, L31 

\bibitem[Kaspi et al.(2003)]{2003ApJ...588L..93K} Kaspi, V.~M., Gavriil, 
F.~P., Woods, P.~M., Jensen, J.~B., Roberts, M.~S.~E., \& Chakrabarty, D.\ 
2003, \apjl, 588, L93 

\bibitem[Kaspi et al.(2006)]{2006ATel..794....1K} Kaspi, V., Dib, R., \& 
Gavriil, F.\ 2006, The Astronomer's Telegram, 794, 1 

\bibitem[Kaspi (2006)]{2006London} Kaspi, V. 2006, in 
"Isolated Neutron Stars: From the Interior to the Surface" eds. S. Zane, R.
Turolla, D. Page; Astrophysics \& Space Science in press (astro-ph/0610304)

\bibitem[Krimm]{Krimm}Krimm, H., et al. 2006, GCN Circular 5581 

\bibitem[Mereghetti \& Stella(1995)]{1995ApJ...442L..17M} Mereghetti, S., 
\& Stella, L.\ 1995, \apjl, 442, L17 

\bibitem[Muno et al.(2006)]{2006ApJ...636L..41M} Muno, M.~P., et al.\ 2006, 
\apjl, 636, L41 

\bibitem[Muno et al.(2006b)]{2006ApJ in press} Muno, M.~P., et al.\ 2006b, 
\apjl, submitted

\bibitem[Muno et al.(2006)]{2006ATel..902....1M} Muno, M., Gaensler, B., 
Clark, J.~S., Portegies Zwart, S., Pooley, D., de Grijs, R., Stevens, I., 
\& Negueruela, I.\ 2006, The Astronomer's Telegram, 902

\bibitem[{\"O}zel(2002)]{2002ApJ...575..397O} {\"O}zel, F.\ 2002, \apj, 
575, 397 

\bibitem[Paczy\'nski (1992)]{1992AcA....42..145P} {\rc Paczy\'nski, B.\ 1992,
Acta  Astronomica, 42, 145 }

\bibitem[Pechenick et al.(1983)]{1983ApJ...274..846P} Pechenick, K.~R., 
Ftaclas, C., \& Cohen, J.~M.\ 1983, \apj, 274, 846 

\bibitem[Rea et al.(2004)]{2004A&A...425L...5R} Rea, N., et al.\ 2004, 
\aap, 425, L5 

\bibitem[Rea et al.(2005)]{2005ApJ...627L.133R} Rea, N., Tiengo, A., 
Mereghetti, S., Israel, G.~L., Zane, S., Turolla, R., \& Stella, L.\ 2005, 
\apjl, 627, L133 

\bibitem[Riffert \& Meszaros(1988)]{1988ApJ...325..207R} Riffert, H., \& 
Meszaros, P.\ 1988, \apj, 325, 207 



\bibitem[Skinner et al. (2006)]{skinner}
      Skinner, S.L.,  Perna, R., \& Zhekov, S.A.\ 2006, \apj, {\rc 653, 587}
 
 \bibitem[Thompson \& Duncan(1995)]{1995MNRAS.275..255T} Thompson, C., \& 
Duncan, R.~C.\ 1995, \mnras, 275, 255 

\bibitem[Tiengo et al.(2005)]{2005A&A...437..997T} Tiengo, A., Mereghetti, 
S., Turolla, R., Zane, S., Rea, N., Stella, L., \& Israel, G.~L.\ 2005, 
\aap, 437, 997 

\bibitem[Wang et al.(2006)]{2006ATel..910....1W} Wang, Z., Kaspi, V.~M., 
Osip, D., Morrell, N., Kaplan, D.~L., \& Chakrabarty, D.\ 2006, The 
Astronomer's Telegram, 910
 
\bibitem[Wong et al.(2001)]{2001ApJ...548..447W} Wong, T., Backer, D.~C., 
\& Lyne, A.~G.\ 2001, \apj, 548, 447 

\bibitem[Woods \& Thompson (2004)]{wot04}
      Woods,  P., \& Thompson, C. 2004, in "Compact Stellar X-ray Sources", eds.
W.H.G. Lewin \&  M. van der Klis, astro-ph/0406133 

\bibitem[Woods et al. (2004)]{2004ApJ...605..378W} Woods, P.~M., et al.\ 
2004, \apj, 605, 378 

\bibitem[Woods et al.(2005)]{2005ApJ...629..985W} Woods, P.~M., et al.\ 
2005, \apj, 629, 985 

\bibitem[Woods et al. (2006)]{2006ATel} Woods, P.~M.,  Kaspi, 
V.~M., Gavriil, F.\ 2006, The Astronomer's Telegram, 929

\end{thebibliography}
\end{document}